\documentclass[ amsmath, amssymb, aps, 12pt]{revtex4-1}

\usepackage{graphicx}% Include figure files
\usepackage{dcolumn}% Align table columns on decimal point
\usepackage{bm}% bold math
\usepackage{hyperref}% add hypertext capabilities
\usepackage{xcolor}

\begin{document}

\title{Generating arbitrary laser beam shapes through \\ phase-mapped designed beam splitting}

\author{Pedro Faleiros Silva}
\author{Sérgio Ricardo Muniz}
%\email{srmuniz@ifsc.usp.br}
\affiliation{
Instituto de Física de São Carlos, Universidade de São Paulo, IFSC-USP\\
Caixa Postal 369, CEP 13560-970, São Carlos, SP, Brazil. \\
srmuniz@ifsc.usp.br -- https://orcid.org/0000-0002-8753-4659
}

\begin{abstract}
\vspace{7mm}
We describe here* a method to generate high-definition arbitrary laser beam shapes and profiles useful to many applications, ranging from optical patterning and lithography to optical trapping of microscopic particles and ultracold atoms. The phase contrast between a binary grating and a targeted intensity distribution is encoded on a spatial light modulator to control light diffraction, producing very sharp, speckles-free, and smooth images. Besides simplicity, not requiring additional phase-plates, the method provides straightforward encoding of images onto phase-only masks by a direct pixel mapping, allowing simpler feedback schemes to correct and control light distributions and optical potentials in real-time. \\
(*) \emph{Paper presented at the Conference \href{https://doi.org/10.1109/SBFotonIOPC50774.2021.9461866}{SBFoton-IOPC-2021}.}
\end{abstract}
\maketitle

\section{Introduction}

Producing and precisely controlling the shape of laser beams is essential in many industries and research areas. Wavefront engineering is one of the methods available to create arbitrary patterns with high fidelity. These precisely controlled light patterns can be used in a wide variety of areas \cite{Dickey2017}, from laser cutting and optical 3D printing to basic research in optogenetics and optical trapping experiments.

In optical trapping experiments, like cold atoms and nanotweezers \cite{Marago2013}, fine control over the potential energy landscape \cite{Grimm2000} is key in designing new experiments, especially for ultracold quantum gases and quantum simulations \cite{Ramanathan2011, Ketterer2012, Bernien2017,Eigen2018}. 
The possibility to engineer dynamic and real-time feedback control over the optical potentials is becoming increasingly important to avail new applications in quantum technologies \cite{Ramanathan2011,Bernien2017,Eigen2018,Levine2019}.

Currently, most methods can be divided into two major groups: 1) techniques based on direct amplitude control, as using  Digital Micro Mirrors (DMDs) \cite{Liang2010,Gauthier2016} and Acoustic Optical Modulators (AOMs) \cite{Muniz2006,Henderson2009,Ketterer2012};
2) phase controlled techniques, like holography and phase-contrast methods \cite{Grier2003, Pasienski2008, Gaunt2012, Gauthier2016, Banas2014}. Each approach has its merits and drawbacks. 

The first type is easier to implement and typically has a faster dynamical response, allowing direct and straightforward feedback schemes to correct deviations from the targeted distribution. However, it is limited in its ability to control the phase of the light field and can not create certain types of structured light (like Laguerre-Gauss or Bessel beams). Also, it may suffer from overall intensity disturbances due to speckle, diffraction, or mode distortions.

The second type allows creating more general phase-structured and vector beams, with the possibility to engineer 3D structures, rather than being limited to a region near the focal plane. On the other hand, in addition to a much slower control speed, this method can make it quite challenging to implement real-time feedback to correct localized imperfections of the desired light distribution, especially when using iteratively-designed holograms.

This paper presents a simple phase-controlled approach with some desired features from DMDs and AOMs, such as pixel-to-pixel direct mapping and simple encoding of the desired light pattern. This feature simplifies the implementation of real-time feedback and dynamical control of optical potentials while providing a straightforward method of encoding general target images with impressive fidelity and smoothness.

\section{Methods}
\subsection{Designed beam splitting}
\begin{figure}[htbp]
\centerline{\includegraphics[width=12cm]{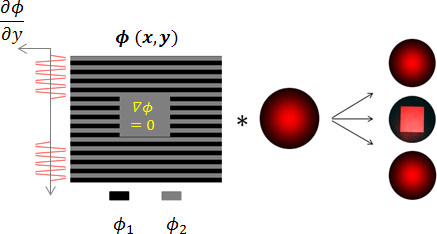}}
\vspace{-5mm}
\caption{Influence of the phase gradient on beam splitting pattern. Regions of constant phase stay at the optical axis (zeroth order of diffraction).
} \label{fig1}
\vspace{2mm}
\end{figure}

A simple example is helpful to understand the main idea here. It is well known that a diffraction grating works by changing the wave vector direction through phase modulation of an impinging wavefront. For example, if a phase $\phi(x,y)$ is applied to a plane wave, the effective wave-vector will be the gradient of the phase, $\Delta \vec{k} = \vec{\nabla}\phi$. Therefore, by spatially controlling the phase gradient is possible to direct coherent light out of its optical axis. This simplified picture does not fully capture the beam diffraction phenomena but gives a valuable intuition on some of its basic principles, as shown in Fig. \ref{fig1}, illustrating a binary phase grating with a central square region of homogeneous phase. 

Regions of constant phase, where $\nabla\phi=0$, will not diffract light out of the beam axis. Therefore, by properly designing  the shape of the phase distribution, one can generate the desired light intensity distribution on the diffracted beams. 

Here, we show experimental results demonstrating that not only simple and flat geometrical shapes can be created using the method described, but also complex and feature-rich images with detailed intensity distributions. 

Our inspiration was the method demonstrated in \cite{Pizolato2007}, which is extended here with a simpler  encoding and different interpretation, producing sharper and smoother results. The new understanding also broadens the range of applications, no longer limited to the zeroth order (on-axis), but, in principle, also applicable to higher orders if properly designed. The essence of the method proposed here is the idea of optimally designed beam splitting through phase imprinting with a spatial light modulator (SLM) and appropriate spatial filtering.
 In simple terms, the method combines a phase mask encoding a targeted image with a simple binary (Ronchi) phase grating to control the shape of the beam splitting, as desired.

\subsection{Phase-mask encoding}

Considering a normalized gray level target image $a(x,y)$, with $N \times M$ pixels, such that $0\leq a(x,y) \leq 1$, a first phase mask is created with phase-relief $e^{ia(x,y)\pi}$ over the full image size. If combined with a second phase mask, composed of intercalated lines with phase-relief set to $e^{i\pi}$, the phase difference $\delta\phi(x,y)=\pi(a(x,y)-1)$, between neighbors pixels along the $y$-direction, it will direct light intensity off the optical axis into the higher diffraction orders, following the designed regions. Therefore, to encode any arbitrary target image, we use the general phase-relief $t(x,y)$:

\begin{equation}
    \begin{split}
        t(x,y) = \sum_{n=0}^{N-1} \{ \text{rect} \left( \frac{y-nP-P/4}{P/2} \right)  e^{ia(x,y)\pi}  \,+ \\ \text{rect} \left( \frac{y-nP-3P/4}{P/2} \right) e^{i\pi} \} \text{rect}(x/s) 
    \end{split}
\label{eq:refname1}
\end{equation}

\begin{figure}[tbhp]
\vspace{+5mm}
\centerline{\includegraphics[width=11cm]{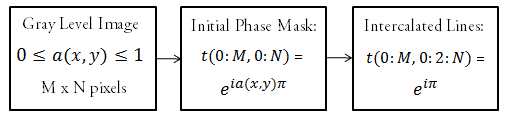}}
\vspace{-5mm}
\caption{Steps to generate the phase mask $t(x,y)$, for any normalized target image $a(x,y)$, of size $M \times N$}
\label{fig2}
\end{figure}

\vspace{5mm}
The diffraction grating period is $P$, and $s$ is the SLM pixel size. Here, we use $P = 2s$. The phase grating is produced by alternating horizontal lines, which generates a vertical phase gradient and beam splitting in that direction. Fig. \ref{fig2} synthesizes the coding procedure.

If important not to lose any image information when applying the  $\pi$-lines encoding, the original image can be duplicated, and the same coding algorithm used.

\begin{figure}[t]
\centerline{\includegraphics[width=14cm]{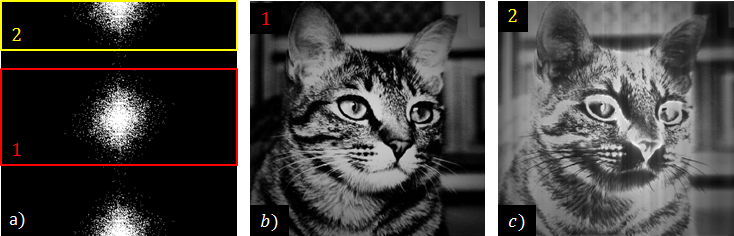}}
\vspace{-3mm}
\caption{Computer simulation of the method. a) Spatial frequency spectrum at the Fourier plane of the encoded beam, showing the diffraction orders $m=0,\pm1$. b) Intensity distribution at the optical axis ($m=0$), and c) at the first orders of diffraction ($m=\pm1$).}
\label{fig3}
\end{figure}

Fig. \ref{fig3} shows a computational simulation of this approach applied to a complex gray-level image. The simulation assumes a Gaussian incident beam with an intensity profile $I_{g}$, and considers a 4f-correlator to form the intensity distribution. This procedure simplifies the diffraction phenomena simulation. First, the phase matrix is created according to Fig. \ref{fig2}. The phase profile is applied at the Gaussian beam electric field modulus ($E_{SLM}=|\sqrt{I_{g}}|e^{it(x,y)}$). Its Fourier Transform is calculated, and the order of diffraction is filtered at the focal plane. Finally, the inverse Fourier transform is calculated and the square modulus is taken to obtain the final image.

The image at Fig. \ref{fig3}b is the final light intensity distribution obtained at the optical axis, after spatial filtering, to keep only the zeroth order of diffraction. It is possible to note the high quality of this method, capable of generating sharp high-definition images, with all the original details. 

Moreover, Fig.\ref{fig3} also sets the idea of this beam splitting method clearly: if the first order of diffraction is selected in spatial filtering, as shown by the yellow rectangle, the final image is the negative of the one observed in the zeroth-order image. This means that one could, alternatively, have the desired image at the (digitally-switchable) first order by using the negative of the desired pattern as the target image.

\section{Experimental Setup}
\begin{figure}[tb]
\centerline{\includegraphics[width=12cm]{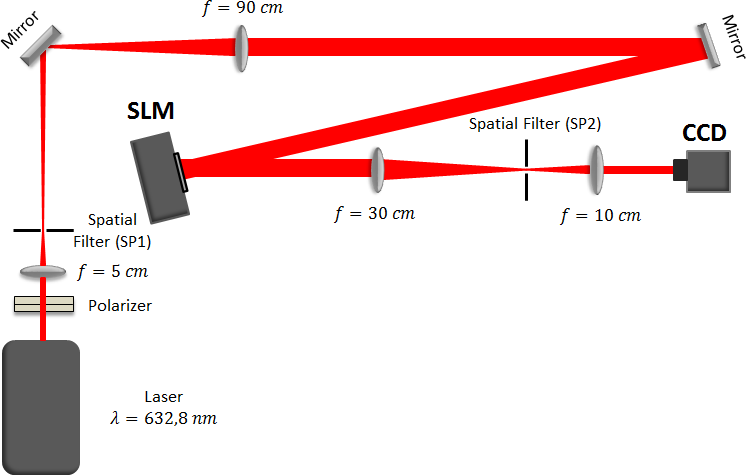}}
\vspace{-5mm}
\caption{Schematic of the optical setup used here.}
\label{fig4}
\end{figure}

The experimental setup for generating and measuring these arbitrary light distributions is illustrated in Fig. \ref{fig4}. A diode laser at $\lambda = 640\,\text{nm}$  is horizontally  polarized to match the orientation of the liquid crystal SLM. The first lens focuses the laser on a $20\,\mu m$ pinhole, working as a first spatial filter (SP1), to get a clean Gaussian beam profile. The second lens collimates back the beam and sets its waist size. The SLM used was a LCOS-SLM (Hamamatsu X13138), with 1272 x 1024 pixels of size $12\,\mu m$, and phase modulation range larger than $2\pi$ at all working wavelenghts. A 4f-correlator was used after the SLM to select the zeroth order mode and control the image's size at the CCD camera. The second spatial filter (SP2) is a simple iris, without small apertures to select the zeroth-order. Here the 4f optical setup is used only for convenience but is not essential since, by Fraunhofer propagation, the zeroth-order could have been selected using other schemes. All measurements (optical images) were taken by a simple CCD camera placed after the 4f-correlator.

\section{Results}
\begin{figure}[htbp]
\centerline{\includegraphics[width=12 cm]{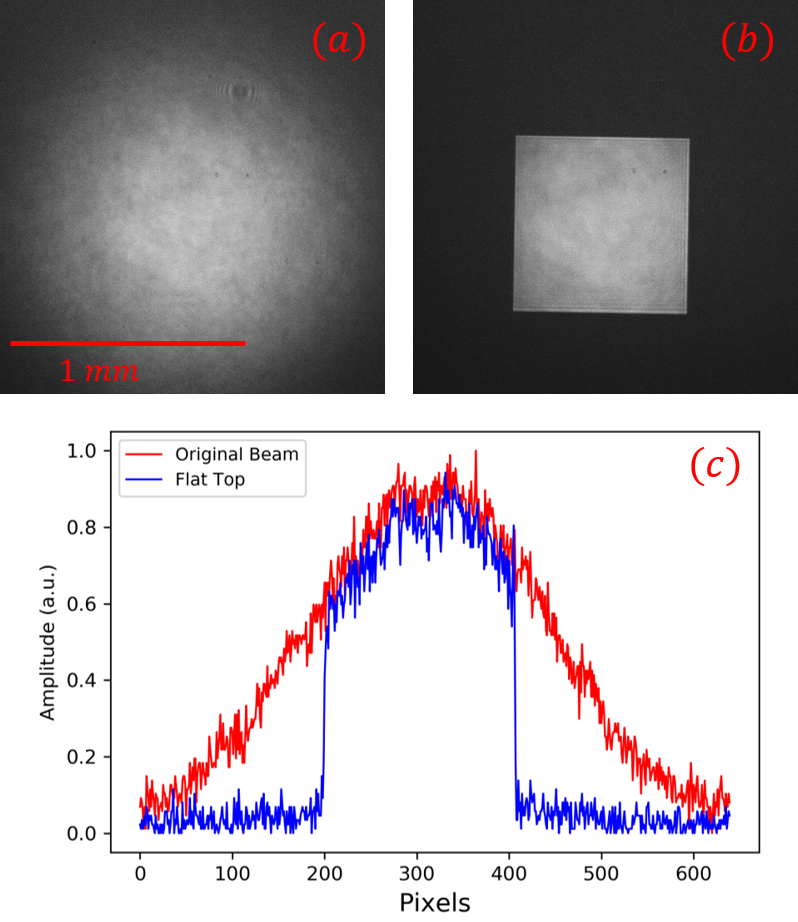}}
\caption{Sharp edge amplitude mask. a) Original Gaussian beam. b) Square shaped optical potential generated by a square flat image target (without intensity correction). c) Gray level 1D intensity profile of both images along the same line. Notice the reltively high intensity contrast at the dark (diffracted) areas. }
\label{fig5}
\end{figure}

\begin{figure}[htbp]
\centerline{\includegraphics[width=12cm]{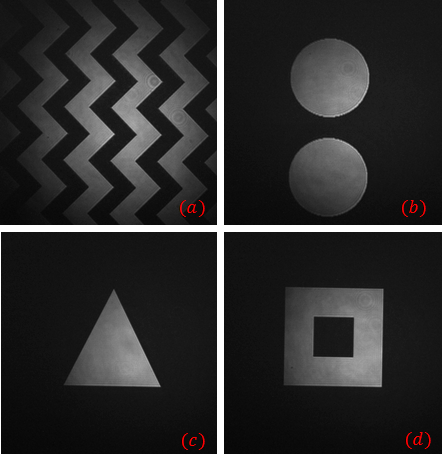}}
\caption{Arbitrary geometrical shapes generated using the method.}
\label{fig6}
\end{figure}

\begin{figure}[htbp]
\centerline{\includegraphics[width=12cm]{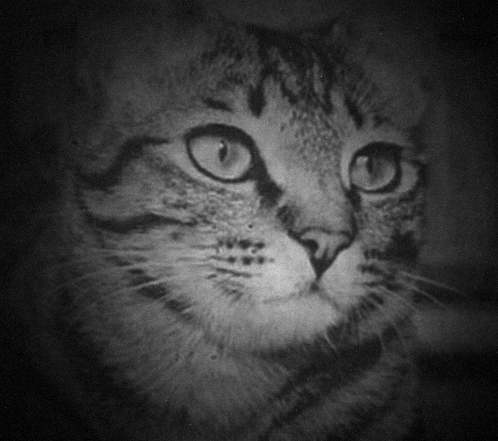}}
\caption{High-definition image obtained experimentally using a feature-rich gray level image as a target. Notice the level o details and absence of added speckle noise or beam distortions.}
\label{fig7}
\end{figure}

Applying the phase encoding procedure described in Fig. \ref{fig2}, we produced arbitrary and high-accuracy optical potentials, as shown in Fig. \ref{fig5}, \ref{fig6}, and \ref{fig7}. Figure \ref{fig5} illustrates an essential feature of this technique: the sharp transitions, which are achievable, demonstrated by the edges of the square shape. Interestingly, the Gaussian profile,  speckles pattern, and other characteristics of the original beam are still present in the bright part of the final image, as expected for a mode preserving beam splitting procedure. We also observed that we can easily flatten the Gaussian intensity profile by applying an inverse Gaussian profile at the target image to get a flat intensity (tophat) beam.

The energy efficiency varies according to the relative size of the target beam pattern compared to the original beam shape and size. 
We empirically observed that the most splitting of the light from the optical axis occurs when a relative phase $\phi_{2}-\phi_{1}=\pi$ is applied to the steps of the binary phase grating. This observation is in agreement with \cite{Zhang1994, Romero2007}.

Figure \ref{fig7} shows the technique's full potential, displaying experimental results of a complex image, full of details, demonstrating excellent fidelity and smoothness. The phase contrast between the image information and the $\pi$-phase lines guides the splitting efficiency at the pixel-level, allowing this kind of high-definition. It is important to note that changes in image information $a(x,y)$ changes directly the final potential, without the necessity of high computational costs or complex optical setups, allowing easier implementation of real-time automated correction algorithms to these optical potentials.

\section{Conclusions}
This paper describes a practical method to generate arbitrarily-shaped, sharp-edge, high-definition optical potentials without added speckle noise or spatial mode distortions, using a direct phase mapping imprinted on the reflected light from an SLM. The straightforward encoding can produce arbitrary beam shapes with smooth and feature-rich intensity control through a pixel-to-pixel phase-mapped designed beam splitting of coherent laser light. The results present quality comparable or better than well-established techniques, with a simple implementation that favors real-time correction of the potentials, applicable to on-axis and off-axis applications.

In principle, this method is not limited to CW lasers, and it could be applied to beam shaping of high-power pulsed lasers and possibly even ultra-fast (fs) lasers for various applications, provided that the damage threshold of the optical components is observed, especially for the SLM. This requirement is not too difficult to achieve, though, as the beam can be expanded to reduce the intensity on more sensitive components. In fact, it is desirable the beam is collimated when reaching the SLM. This is the condition we used in the experiments described here. But, we still did not test this method in other limits, such as diverging or converging beams on the SLM or pulsed lasers. 

If using a fast response SLM, one could use a movie (animated sequence of encoded images) to use this method for time modulation or dynamical control of a beam intensity distribution \cite{Mart2105:Dynamically} or shape, limited only by the bandwidth of the SLM used. The observed diffraction efficiency was typically very high, as one can observe by the beam contrast in Fig. \ref{fig5}, at least if using high-reflectivity SLMs, like the one used in this study.

\section{Acknowledgment}
We acknowledge the financial support provided by FAPESP (Fundação de Amparo à Pesquisa do Estado de São Paulo), the São Paulo Research Foundation, under research Grants  2019/27471-0 and 2013/07276-1, and also the scholarship provided by CAPES.

\bibliographystyle{plain}
\bibliography{References}

\end{document}